\documentclass[pra,onecolumn,showpacs,
10pt]{revtex4-1}
\usepackage[colorlinks,citecolor=blue,linkcolor=red,dvipdfm]{hyperref}
\usepackage{amsmath,amssymb,graphicx}
\usepackage{times}

\begin{document}

\title{Localized modes in nonlinear fractional systems with deep lattices }
\author{Xiuye Liu$^{1,2}$}
\author{Boris A. Malomed$^{3,4}$}
\author{Jianhua Zeng$^{1,2}$}
\affiliation{$^1$State Key Laboratory of Transient Optics and Photonics, Xi'an
Institute of Optics and Precision Mechanics of CAS, Xi'an 710119, China}
\affiliation{University of Chinese Academy of Sciences, Beijing 100049,
China}
\affiliation{$^3$Department of Physical Electronics, School of Electrical
Engineering, Faculty of Engineering, and the Center for Light-Matter
Interaction, Tel Aviv University, Ramat Aviv, Tel Aviv P.O.B. 39040, Israel}
\affiliation{$^4$Instituto de Alta Investigaci\'{o}n, Universidad de
Tarapac\'{a}, Casilla 7D, Arica, Chile}
\affiliation{$^2$State Key Laboratory of Transient Optics and Photonics, Xi'an
Institute of Optics and Precision Mechanics of CAS, Xi'an 710119, China}
\affiliation{University of Chinese Academy of Sciences, Beijing 100049,
China}

\begin{abstract}
Solitons in the fractional space, supported by lattice potentials, have
recently attracted much interest. We consider the limit of deep one- and
two-dimensional (1D and 2D) lattices in this system, featuring finite
bandgaps separated by nearly flat Bloch bands. Such spectra are also a
subject of great interest in current studies. The existence, shapes, and
stability of various localized modes, including fundamental gap and vortex
solitons, are investigated by means of numerical methods; some results are
also obtained with the help of analytical approximations. In particular, the
1D and 2D gap solitons, belonging to the first and second finite bandgaps,
are tightly confined around a single cell of the deep lattice. Vortex gap
solitons are constructed as four-peak \textquotedblleft squares" and
\textquotedblleft rhombuses" with imprinted winding number $S=1$. Stability
of the solitons is explored by means of the linearization and verified by
direct simulations.
\end{abstract}

\maketitle

\section{Introduction}

{\label{Sec:1}} ~\newline

Fractional calculus, different forms of which have been elaborated in
mathematics, is, essentially, a theory of fractional differentiation and
integration~\cite{FC-book0,FC-book1,FC-book2,FC-book3,FC-book4}. More
recently, it has found applications to modeling various phenomena in quantum
mechanics ~\cite{Lask1,Lask2,Lask3}, optics~\cite{Fujioka,Frac-optics},
ultracold atomic gases~\cite{frac-BEC-diffusion-E,frac-BEC-diffusion-T}, and
condensed matter~\cite{Frac-condensed,Frac-polariton}. In particular, the
formulation of quantum mechanics based on Feynman path integrals, if applied
to particles moving by stochastic Levy flights, instead of the usual
Brownian motion, leads to the fractional Schr\"{o}dinger equation~\cite%
{Lask1,Lask2,Lask3}, see also a recent book \cite{Frac-book} summarizing
this theory. \vspace{5mm}

In the framework of these studies, the fractional Schr\"{o}dinger equation
and its extensions in the form of nonlinear fractional Schr\"{o}dinger
equations (NLFSEs) have drawn a great deal of interest, revealing a variety
of linear and nonlinear wave patterns in diverse physical media \cite%
{Frac1,Frac2,Frac3,Frac3b,Frac4,Frac5,Frac5b,Frac6,Frac7,Frac7b,Dong,Frac8,Frac9,Frac10,Frac11,Frac12,Frac-CQ1d,Frac-CQ2d,Frac-SSB,Frac-saturable,Frac-coupler,Frac-SSBLinear,Frac-CQ,Frac-PRR,Frac-CGL,Frac-PTOL,Pengfei}%
, see also a recent brief review \cite{review}. These include gap solitons
(GSs) trapped by shallow or moderately deep lattice potentials acting in the
combination with the self-repulsive nonlinearity \cite%
{Frac6,Frac7,Frac7b,Dong,Frac12,Frac-CQ1d,Frac-CQ2d,Frac-GS-PT,Frac-PTOL},
while the deep-lattice limit was not studied in detail. It is relevant to
mention that optical and matter-wave GSs, supported by the interplay of
periodic potentials, self-repulsion, and normal (non-fractional) diffraction
\cite%
{FPC,RMP06,soliton-periodic,PC,NL-RMP,PL,NRP,BEC-darkGap,NL-focus,BEC-darkGapQ,BEC-EIT1d,BEC-EIT2d}%
, have been experimentally created in fiber Bragg gratings and photonic
crystals~\cite{GS-FBG,GS-WA,moving-gap-sol,GS-HPL}, in atomic Bose-Einstein
condensates loaded into optical lattices (OLs) \cite{GS-BEC}, and in
polariton condensates trapped in semiconductor microcavities~\cite%
{GS-EPC1,GS-EPC2,GS-EPC3}. Recently, novel periodic potentials, such as moir%
\'{e} lattices \cite{moire1,moire2}, which feature flat-band spectra,
similar to those induced by deep lattices, were introduced in this context.
\vspace{5mm}

The nonlinear Schr\"{o}dinger equation with normal (non-fractional)
diffraction and a deep lattice potential is often replaced, in the
tight-binding approximation, by its discrete version with the
nearest-neighbor coupling. The accuracy of this limit is confined to the
first finite bandgap \cite{Smerzi}, and GSs correspond to discrete solitons
with the staggered structure \cite{staggered}. However, to address GSs in
fractional models including deep OL potentials, which is the subject of the
present work, it is necessary to deal with NLFSEs in the continuous form, as
the fractional diffraction is represented by a nonlocal integral operator.
Recently, a discrete approximation for the one- and two-dimensional (1D and
2D) NLFSE with the self-attractive nonlinearity was introduced in Refs. \cite%
{Discrete1d,Discrete2d,FDNLS-vortex}. It takes the form of a nonlocal
discrete equation with a complex form of the long-range interaction, which
produces discrete solitons. However, in the case of self-repulsion, the
staggering transformation does not apply to discrete equations in which the
interaction cannot be limited to the nearest neighbors.\vspace{5mm}

In this work, we address, by means of numerical and analytical methods, the
existence and stability of various localized modes, including 1D and 2D GSs
and gap-vortex modes (cf. Ref. \cite{HS}) in 2D, in the first and second
finite bandgaps. In particular, the GSs are always found as highly confined
modes nearly squeezed into a single lattice cell, even if they are excited
in the second bandgap, which the discrete limit would fail to grasp. On the
contrary, GSs supported by shallow or moderately deep lattice potentials
normally cover several (or many) cells. The same is true for discrete
solitons, including ones in the fractional model with the self-attraction
\cite{Discrete1d}. Stability of the solitons in the system under the
consideration is explored by means of the linear-stability analysis and
direct simulations. \vspace{5mm}

\begin{figure}[t]
\begin{center}
\includegraphics[width=0.7\columnwidth]{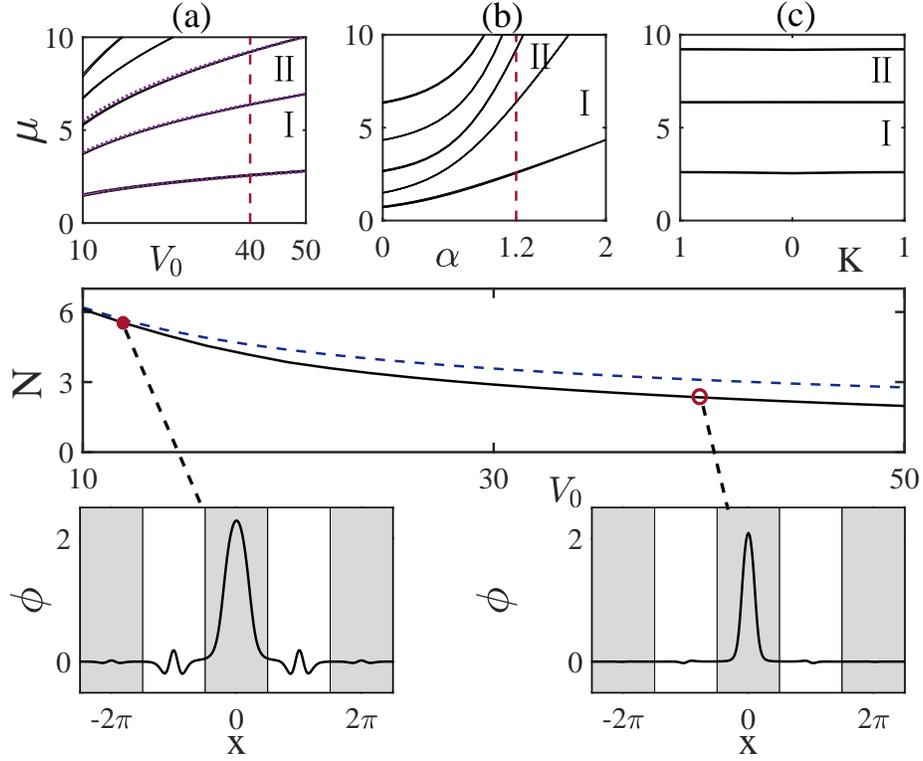}
\end{center}
\caption{Linear bandgap spectra, shown by values of chemical potential $%
\protect\mu $ vs. the lattice depth $V_{0}$ (a), LI (L\'{e}vy index) $%
\protect\alpha $ (b), and quasi-momentum $K$ (c) in the 1D model. Other
parameters are fixed as $\protect\alpha =1.2$ in (a,c) and $V_{0}=40$ in
(b,c) (these values are designated by vertical dashed lines in (b) and (a),
respectively). Here and below in Figs. \protect\ref{fig2}$\sim$ \protect\ref{fig5}, regions I and II represent, respectively, the first and second finite bandgaps. Purple dotted lines in the first three Bloch bands in (a) represent the approximate relation given by Eq. (\protect\ref{mu}), i.e., $\protect\mu =CV_{0}^{3/8}$
for $\protect\alpha =1.2$, where the fitting constants are $C=0.64$, $1.60$,
and $2.31$ for the three lines. The middle plot: norm $N$ of the GSs (gap
solitons) in the first finite bandgap supported by the deep OL is shown vs. $%
V_{0}$ by the continuous line, for $\protect\mu =6$ and $\protect\alpha =1.2$%
. The blue dashed line shows the same dependence as predicted by the TF
(Thomas-Fermi) approximation. Typical profiles of 1D GSs in the first finite
bandgap are displayed by the bottom plots at $V_{0}=12$ (left, $N=5.42$) and
$V_{0}=40$ (right, $N=2.34$). Here and in similar figures below, alternating
gray and white stripes in the bottom panels designate periods of the lattice
potential.}
\label{fig1}
\end{figure}

\section{The Model}

{\label{Sec:2}} \vspace{5mm}

Our setting is based on the continuous NLFSE of the Gross-Pitaevskii type
\cite{Pit}, for a mean-field wave function $U(X,Y,T)$ of particles moving by
L\'{e}vy flights. With time $T$ and coordinates $\left( X,Y\right) $
measured in physical units, the equation is written as
\begin{equation}
i\hbar \frac{\partial U}{\partial T}=D_{\alpha }\left( -\hbar ^{2}\nabla
^{2}\right) ^{\alpha /2}U+W_{\mathrm{OL}}(X,Y)U+\frac{2\sqrt{2\pi }\hbar
^{2}a_{s}}{ma_{\perp }}\left\vert U\right\vert ^{2}U,  \label{GP}
\end{equation}%
where $D_{\alpha }$ is the Laskin's coefficient \cite{Lask3} with dimension
\begin{equation}
\left[ D_{\alpha }\right] =\left[ \mathrm{mass}\right] ^{1-\alpha }\left[
\mathrm{length}\right] ^{2-\alpha }\left[ \mathrm{time}\right] ^{-\left(
2-\alpha \right) },  \label{dim}
\end{equation}%
which carries over into the usual one, $\left( 2m\right) ^{-1}$, in the
integer limit ($\alpha =2$), $W_{\mathrm{OL}}\left( X,Y\right) $ is the
spatially periodic potential induced by the OL, $a_{s}>0$ is the scattering
length, $m$ the atomic mass, and $a_{\perp }\equiv \sqrt{\hbar /\left(
m\Omega \right) }$ is the confinement size imposed in the third direction, $Z
$, by the usual trapping potential, $W_{\mathrm{trap}}(Z)=\left( m/2\right)
\Omega ^{2}Z^{2}$ \cite{Salasnich}. The fractional kinetic-energy operator
in Eq. (\ref{GP}) is determined by means of the direct and inverse
Fourier-transform operators, $\mathcal{F}$ and $\mathcal{F}^{-1}$, following
the concept of the fractional \textit{Riesz derivative} \cite{Riesz}:%
\begin{equation}
\left( -\nabla ^{2}\right) ^{{\alpha }/{2}}f\left( \mathbf{R}\right) \equiv
\mathcal{F}^{-1}\left[ \left\vert \mathbf{K}\right\vert ^{\mathrm{\alpha }}%
\mathcal{F}\left( f\right) \right] =\frac{1}{\left( 2\pi \right) ^{2}}\int
\int {\left\vert \mathbf{K}\right\vert ^{\mathrm{\alpha }}\hat{f}\left(
\mathbf{K}\right) \exp }\left( i\mathbf{K\cdot R}\right) {d\mathbf{K}}.
\label{fractional}
\end{equation}%
Here $f(\mathbf{R})$ is a arbitrary function of coordinates $\left(
X,Y\right) $, $f(\mathbf{K})$ is its Fourier transform, and $\alpha $ is the
L\'{e}vy index (LI), taking values $0<\alpha \leq 2$. In the 1D case, the
corresponding operator $\left( -\partial ^{2}/\partial x^{2}\right) ^{\alpha
/2}$ is defined by the straightforward counterpart of Eq. (\ref{fractional}%
). In 1D systems with the attractive cubic nonlinearity, which corresponds
to $a_{s}<0$, interval $0<\alpha \leq 1$ is not considered, as the collapse
occurs in it, and in the same 2D setting all values $\alpha \leq 2$ lead to
the collapse for $a_{s}<0$ \cite{review}). However, we here consider Eq. (%
\ref{GP}) with the self-defocusing sign of the nonlinearity, $a_{s}>0$.%
\vspace{5mm}

It is relevant to mention that the mathematical formalism was developed for
different definitions of fractional derivatives. In particular, while the
Riesz form (\ref{fractional}) corresponds to the implementation in quantum
mechanics \cite{Frac-book} and optics \cite{Frac-optics}, the above-mentioned
works \cite{Discrete1d,Discrete2d,FDNLS-vortex}, which aim to introduce the
discrete version of the NLFSE, adopted a different, Riemann-Liouville,
definition of the fractional differentiation. It is expected the different
definitions lead to similar but not identical results.\vspace{5mm}

Equation (\ref{GP}) can be made dimensionless by setting%
\begin{eqnarray}
\left( X,Y\right) &\equiv &\frac{d}{\pi }\left( x,y\right) ,T\equiv \left(
\frac{d}{\pi }\right) ^{\alpha }\frac{\hbar ^{1-\alpha }}{2D_{\alpha }}t,W_{%
\mathrm{OL}}\equiv D_{\alpha }\left( \frac{\pi \hbar }{d}\right) ^{\alpha
}V_{\mathrm{OL}},  \notag \\
U\left( X,Y,T\right) &\equiv &U_{0}\psi \left( x,y,t\right) ,U_{0}=\frac{\pi
^{(1/2)\left( \alpha -1/2\right) }}{2^{1/4}d^{\alpha /2}\hbar ^{1-\alpha /2}}%
\sqrt{\frac{D_{\alpha }ma_{\perp }}{a_{s}}}.  \label{scaling}
\end{eqnarray}%
The accordingly rescaled Eq. (\ref{GP}) is
\begin{equation}
i\frac{\partial \psi }{{\partial t}}=\frac{1}{2}\left( {-\nabla ^{2}}\right)
^{\alpha /2}\psi +V_{\mathrm{OL}}(x,y)\psi +\left\vert \psi \right\vert
^{2}\psi .  \label{model}
\end{equation}%
The number of atoms in the condensate, $\mathcal{N}$, whose evolution is
governed by Eq. (\ref{model}) is expressed in terms of norm $N$ of the
scaled wave function as per Eq. (\ref{scaling}):
\begin{equation}
\mathcal{N}=\left( U_{0}d/\pi \right) ^{2}N,  \label{number}
\end{equation}%
\begin{equation}
N\equiv \int \int \left\vert \psi \left( x,y\right) \right\vert ^{2}dxdy.
\label{N}
\end{equation}%
Taking into regard Eq. (\ref{dim}), it is easy to check that expression (\ref%
{number}) is dimensionless, as it should be. For the condensate of $^{87}$Rb
atoms, loaded in the OL potential with period $d=500$ nm, Eq. (\ref{scaling}%
) shows that $t=1$ corresponds, in physical units, to $\simeq 0.04$ ms, and,
with the transverse-confinement size $a_{\perp }\simeq 5$ $\mathrm{\mu }$m, $%
N=1$ represents $\simeq 300$ atoms. Similar rescalings are available in the
1D case, with the single coordinate, $x$. \vspace{5mm}

Previously reported experimental observations of effectively one-dimensional
GSs in the condensate of $^{87}$Rb demonstrated that the soliton was built
of ca. $250$ atoms ~\cite{GS-BEC}. More extended \textquotedblleft gap
waves" were observed as chains of several closely overlapped GSs, containing
$\simeq 5000$ atoms~\cite{GW-BEC}. Thus, the solitons predicted here should
be within the reach of the available experimental techniques. \vspace{5mm}

The scaled OL potential with lattice depth $V_{0}$ is taken as
\begin{equation}
V_{\mathrm{OL}}(x)=V_{0}\sin ^{2}x  \label{1D}
\end{equation}%
in the 1D space, or
\begin{equation}
V_{\mathrm{OL}}(x,y)=V_{0}\left( \sin ^{2}x+\sin ^{2}y\right) ,  \label{2D}
\end{equation}%
in 2D (in the latter case, the full depth is $2V_{0}$). The self-repulsive
nonlinearity in Eq. (\ref{model}) is expected to create GS modes populating
finite bandgaps of the linear Bloch-wave spectrum induced by the OL. In
terms of optics, Eq. (\ref{model}), with $t$ replaced by the scaled
propagation distance, applies to the spatial-domain propagation of light in
media with the effective fractional diffraction induced by means of the
scheme which converts the spatial optical beam into the Fourier space and
applies an appropriate phase shift in it, then getting back into the spatial
domain \cite{Frac-optics}. In that case, norm (\ref{N}) is the scaled power
of the optical beam, and the deep spatially periodic potential can be
realized in a photonic crystal composed of alternating circular or planar
layers (in the 2D and 1D cases, respectively) with high and low values of
the refractive index \cite{PhotCryst}. \vspace{5mm}

We search for soliton solutions with chemical potential $\mu $ in the usual
form, $\psi =\phi \left( x,y\right) e^{-i\mu t}$. Substituting this in Eq. (%
\ref{model}) leads to the equation for the stationary wave function:
\begin{equation}
\mu \phi =\frac{1}{2}\left( {-\nabla ^{2}}\right) ^{\alpha /2}\phi +V_{%
\mathrm{OL}}(x,y)\phi +\left\vert \phi \right\vert ^{2}\phi .
\label{station}
\end{equation}

Once stationary soliton solutions $\phi (x,y)$ are obtained by solving Eq. (%
\ref{station}), it is then necessary to check their stability. For this
purpose, we apply the linear-stability analysis method which, as usual, is
done by taking the perturbed wave function,\newline
\begin{equation}
\psi \left( {x,y,t}\right) =e^{-i\mu t}\left[ {\phi \left( x,y\right)
+u\left( x,y\right) e^{\lambda t}+v^{\ast }\left( x,y\right) e^{\lambda
^{\ast }t}}\right] ,  \label{pert}
\end{equation}%
with undisturbed field amplitude $\phi (x,y)$, small perturbations $u(x,y)$
and $v^{\ast }(x,y)$, and eigenvalue $\lambda $, the asterisk standing for
the complex conjugate. The substitution of ansatz (\ref{pert}) into Eq. (\ref%
{model}) and linearization with respect to the small perturbation leads to
the eigenvalue problem based on the following equations:
\begin{equation}
{i\lambda u\mathrm{\ =+}\mu u+\frac{1}{2}\left( {-\nabla ^{2}}\right)
^{\alpha /2}u+\phi ^{2}(2u+v)}+V_{\mathrm{OL}}u,  \label{LSA1}
\end{equation}%
\begin{equation}
{i\lambda v=-\mu v-\frac{1}{2}\left( {-\nabla ^{2}}\right) ^{\alpha
/2}v-\phi ^{2}(2v+u)}-V_{\mathrm{OL}}v.  \label{LSA2}
\end{equation}%
The underlying solution is stable if all eigenvalues produced by numerical
solution of Eqs. (\ref{LSA1}) and (\ref{LSA2})] have zero real parts, $%
\lambda _{R}=0$. \vspace{5mm}

The numerical analysis presented below first produces solutions $\phi (x,y)$
of Eq. (\ref{station}), using the modified squared-operator method \cite%
{MSOM}. Then, their linear stability is explored for eigenmodes of small
perturbations obtained as solutions of Eqs. (\ref{LSA1}) and (\ref{LSA2}).
Finally, the dynamical stability is tested in direct simulations of Eq. (\ref%
{model}) employing the fourth-order Runge-Kutta method based on the fast
Fourier transform.

\begin{figure}[t]
\begin{center}
\includegraphics[width=0.7\columnwidth]{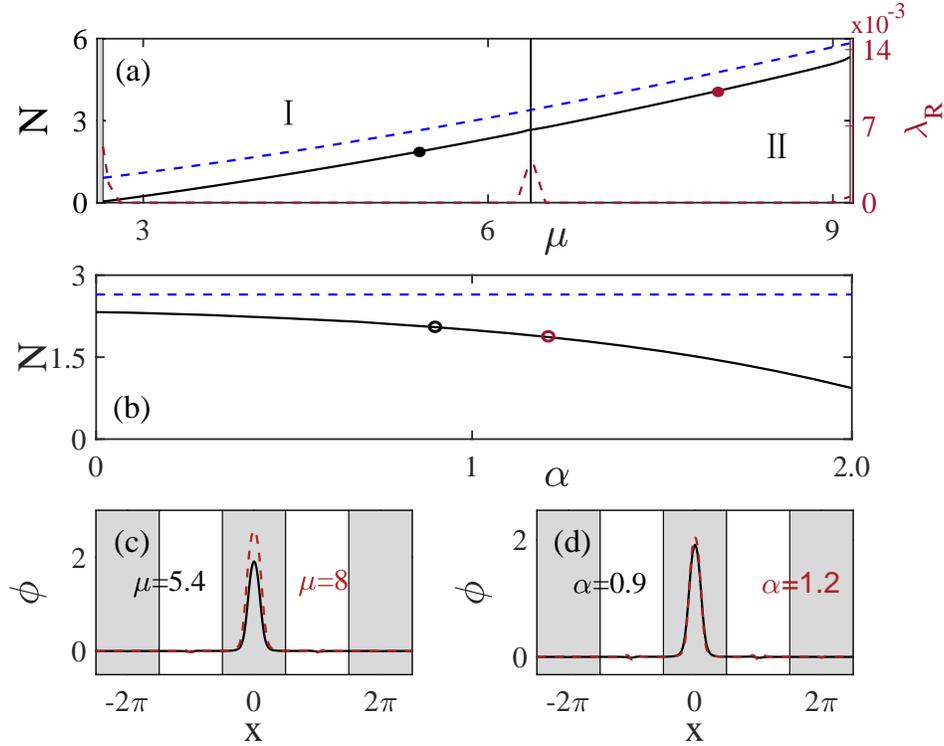}
\end{center}
\caption{Norm $N$ as a function of chemical potential $\protect\mu $ (a) and
LI $\protect\alpha $ (b) for 1D GSs in the fractional nonlinear system with
the deep OL potential (the depth is fixed as $V_{0}=40$). To quantify
stability of the solitons, the largest real part of the perturbation growth
rate, $\protect\lambda $, is shown as a function of $\protect\mu $ by the
red dashed line in (a). Other parameters are $\protect\alpha =1.2$ in (a)
and $\protect\mu =4$ in (b). Panel (a) shows the results for the GSs in the
two lowest bandgaps, separated by the flat Bloch band at $\protect\mu %
\approx 6.37$. The result produced by the TF (Thomas-Fermi) approximation is
shown by the blue dashed line in (a). The TF-predicted value of $N$, which
does not depend on $\protect\alpha $, is shown as a reference one by the
horizontal blue dashed line in (b). Profiles of GSs with different values of
$\protect\mu $ (c) and $\protect\alpha $ (d), which correspond,
respectively, to the marked points in (a) and marked circles in (b), are
displayed in the bottom panels.}
\label{fig2}
\end{figure}

\section{The one-dimensional bandgap spectrum and gap solitons (GSs)}

\vspace{5mm}

To explore the nonlinear dynamics of GSs trapped in the lattice potential,
the bandgap spectrum of the underlying linearized equation for the
Floquet-Bloch modes is required. They are defined as usual:%
\begin{equation}
\phi _{\mathrm{K}}(x)=\Phi _{\mathrm{K}}(x)\exp (i\mathrm{K}x),  \label{K}
\end{equation}%
where $\mathrm{K}$ is the quasi-momentum, and $\Phi (x)$ is a periodic
function \cite{FPC,RMP06,soliton-periodic,PC}. The spectrum was produced by
a numerical solution of the linearized version of Eq. (\ref{station}). In
Fig. \ref{fig1}(a), the 1D bandgap spectrum is plotted for the OL potential (%
\ref{1D}), varying lattice depth $V_{0}$ at a fixed LI, $\alpha =1.2$. It is
seen that widths of the first two finite bandgaps stay nearly constant as $%
V_{0}$ varies in the interval of $[10,30]$. At a fixed value of $V_{0}$ from
this interval, the bandgap widens with the increase of $\alpha $, as shown
in Fig. \ref{fig1}(b). The latter feature is similar to its counterpart
found in fractional systems with shallow OLs \cite{Frac6,Dong,Frac-CQ1d}.
\vspace{5mm}

Figure \ref{fig1}(c) shows the bandgap spectrum vs. quasimomentum $\mathrm{K}
$ for $\alpha =1.2$ and $V_{0}=40$. Note that the Bloch bands are very flat
for the deep lattice, because the tunnel coupling between adjacent potential
wells is weak. The flat-band phenomenology, including the wave propagation
and formation of solitons, has been recently developed in models based on
the usual nonlinear Schr\"{o}dinger equation ($\alpha =2$) with other
periodic potentials \cite{flat0}-\cite{flat}, including recent works with
moir\'{e} lattices~\cite{moire1,moire2}.\vspace{5mm}

The nearly constant values of $\mu $ corresponding to the flat bands cannot
be found analytically. However, it is possible to predict their dependence
on the potential's depth, $V_{0}$. Indeed, individual deep wells in lattice
potentials (\ref{1D}) and (\ref{2D}) may be approximated by
harmonic-oscillator (HO) potentials, \textit{viz}.,%
\begin{equation}
V_{\mathrm{HO}}^{\mathrm{(1D)}}(x)\approx V_{0}x^{2}  \label{1DOH}
\end{equation}%
in the 1D case, and
\begin{equation}
V_{\mathrm{HO}}^{\mathrm{(2D)}}(x,y)\approx V_{0}\left( x^{2}+y^{2}\right)
\label{2DOH}
\end{equation}%
in 2D. Then, values of $\mu $ which determine the flat bands may be
approximately found as eigenvalues of the linearized version of Eq. (\ref%
{station}) with the HO potential (\ref{1DOH}) or (\ref{2DOH}). This problem
does not admit an exact solution in the fractional space \cite{no exact
solution}. Nevertheless, it is easy to predict an exact scaling in the
dependence of these eigenvalues on the potential's strength, $V_{0}$ (it is
the same for the 1D and 2D settings):%
\begin{equation}
\mu _{\mathrm{flat-band}}\sim V_{0}^{\alpha /\left( 2+\alpha \right) }.
\label{mu}
\end{equation}%
Simultaneously, the size of the wave function trapped in potential (\ref%
{1DOH}) or (\ref{2DOH}) scales as $x_{0}\sim V_{0}^{-1/\left( 2+\alpha
\right) }$. In the case of $\alpha =2$, these scaling relations correspond
to the ones commonly known for the HO in quantum mechanics, $\mu _{\mathrm{%
flat-band}}\sim \sqrt{V_{0}}$ and $x_{0}\sim V_{0}^{-1/4}$ (the latter one
represents the standard HO length). In Fig. \ref{fig1}(a), we have plotted
the dependence predicted by Eq. (\ref{mu}) for the first three Bloch bands,
showing a very close match. \vspace{5mm}

Usually, in deep OLs localized nonlinear modes occupy, essentially, a single
cell of the lattice (which suggests a possibility of approximating them by
the discrete lattice models, as mentioned above) \cite{Smerzi,L-NL1d}. This
property is corroborated by the modal profiles displayed in the third line
of Fig. \ref{fig1}. The GSs shown in it for $\mu =6$, $\alpha =1.2$ belong
to the first finite bandgap, as per Fig. \ref{fig1}(a). At a moderate
lattice depth, $V_{0}=12$, the GS features undulating tails, while a very
deep lattice, with $V_{0}=40$ produces a GS with a compact shape strongly
confined to a single cell, which is a typical feature of localized modes
sustained in deep lattices. A simple asymptotic consideration of Eq. (\ref%
{station}) demonstrates that, at large $|x|$, the tails decay as%
\begin{equation}
\phi ^{2}(x)\sim \exp \left( -\mathrm{const}\cdot V_{0}^{1/\alpha
}|x|^{-\left( 1+2/\alpha \right) }\right) ,  \label{tail}
\end{equation}%
i.e., as a super-Gaussian at $\alpha <2$. In the 2D case, the same
asymptotic expression (\ref{tail}) is valid, with $|x|$ replaced by radial
coordinate $r$.

Hereafter, we set $V_{0}=40$ [as marked by the red dashed line in Fig. \ref%
{fig1}(a)] and $\alpha =1.2$ [marked by the red dashed line in Fig. \ref%
{fig1}(b)] for the study of the localized modes in the NLFSE with the 1D
deep lattice.

\begin{figure}[t]
\begin{center}
\includegraphics[width=0.7\columnwidth]{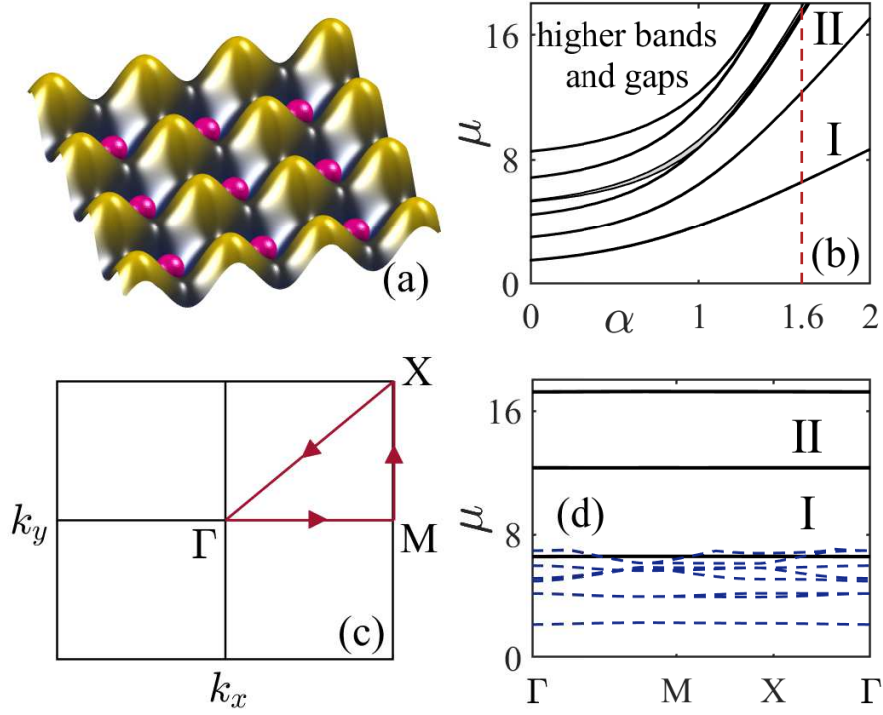}
\end{center}
\caption{(a) The stereographic representation of the 2D optical lattice
potential, at the bottom of which bosonic atoms are trapped, in real space.
(c) The corresponding first reduced Brillouin zone in the reciprocal lattice
space; X, M, and $\Gamma$ denote the high-symmetry points in the irreducible zone. (b) Dependences of chemical potential $\protect\mu$ on L\'{e}vy index
$\protect\alpha $, and (d) on Bloch quasi-momentum $\mathrm{K}$, at $\protect%
\alpha =1.6$ [this value is designated by the vertical dashed red line in
(b)]. In panels (b) and (d) the dependences are plotted for lowest Bloch
bands. Higher-order bands and gaps between them, which occupy the top left
corner in (a) (as marked in that panel) are not shown here in detail. Here
and below, the lattice depth is $V_{0}=40$. The blue dashed lines in panel
(d) correspond to the linear-Bloch spectrum for a shallow lattice with $%
V_{0}=4$.}
\label{fig3}
\end{figure}

\vspace{5mm}

Dependences of norm $N$ on chemical potential $\mu $ and LI $\alpha $ for
the family of 1D GSs sustained by the self-defocusing nonlinearity and deep
OL in the two lowest bandgaps of the fractional system are shown in Figs. %
\ref{fig2}(a) and (b), respectively. One can see from the Fig. \ref{fig2}(a)
that dependence $N(\mu )$ agrees with the necessary stability condition for
solitons in settings with repulsive nonlinearities, \textit{viz}., the
\textquotedblleft anti-Vakhitov--Kolokolov" (anti-VK) criterion, $dN/d\mu >0$%
~\cite{L-NL1d}, which was tested in many other models \cite%
{L-NL2d,GS-PT1,Frac-CQ1d,BEC-darkGap,BEC-darkGapQ} (the VK criterion per se,
$dN/d\mu <0$, applies to systems with self-attraction \cite{VK,Berge,Fibich}%
). Also displayed in Fig. \ref{fig2}(a) are results of the linear-stability
analysis, which demonstrates that the GSs are almost completely stable in
the two lowest finite bandgaps, narrow instability intervals occurring at
top edges of both bandgaps. Two typical examples of stable GSs residing in
these bandgaps are presented in Fig. \ref{fig2}(c). These examples clearly
confirm that the GSs keep an almost tailless shape, in accordance with Eq. (%
\ref{tail}), being effectively confined in a single OL cell. On the
contrary, GSs in nonlinear fractional systems with a shallow lattice
potential are usually broad modes extending over several lattice cells \cite%
{Frac-CQ1d,Frac-CQ2d}. For the GSs with fixed $\mu $, the increase of LI
results in a decrease of the norm necessary for the existence of the
solitons, as seen in Fig. \ref{fig2}(b).

\vspace{5mm}

The fact that the solitons are tightly confined in the single potential cell
makes it natural to compare their shapes to the prediction of the
Thomas-Fermi (TF) approximation, which neglects the diffraction term in Eq. (%
\ref{station}) \cite{Pit}:

\begin{equation}
\phi _{\mathrm{TF}}^{2}(x)=\left\{
\begin{array}{c}
\mu -V_{0}\sin ^{2}x,~\mathrm{at}~|x|<\arcsin \left( \sqrt{\mu /V_{0}}%
\right) , \\
0,~\mathrm{at}~|x|>\arcsin \left( \sqrt{\mu /V_{0}}\right) .%
\end{array}%
\right.  \label{TF-profile}
\end{equation}%
The respective norm is given by%
\begin{equation}
N_{\mathrm{TF}}=2\int_{0}^{\arcsin \left( \sqrt{\mu /V_{0}}\right) }\phi _{%
\mathrm{TF}}^{2}(x)dx=\sqrt{\mu \left( V_{0}-\mu \right) }-\left( V_{0}-2\mu
\right) \arcsin \left( \sqrt{\frac{\mu }{V_{0}}}\right) .  \label{NTF}
\end{equation}%
For the case of the deep OL, a practically important particular case is the
one with $\mu \ll V_{0}$. In this case, expression (\ref{NTF}) simplifies to

\begin{equation}
N_{\mathrm{TF}}\approx 4\mu ^{3/2}/\left( 3\sqrt{V_{0}}\right) .
\label{TFapprox}
\end{equation}
\vspace{5mm}

The TF predictions produced by Eqs. (\ref{NTF}) and (\ref{TFapprox}) are
displayed by the blue dashed lines in Figs. \ref{fig2}(a, b). In fact, the
numerical findings are most interesting in the case when they are \emph{%
conspicuously different} from the TF approximation, i.e., at $\mu $ not too
large, and at $\alpha $ not too close to $0$, as the proximity to the TF
approximation, which ignores the fractional diffraction, implies that it is
not an essential factor.

\begin{figure}[t]
\begin{center}
\includegraphics[width=0.7\columnwidth]{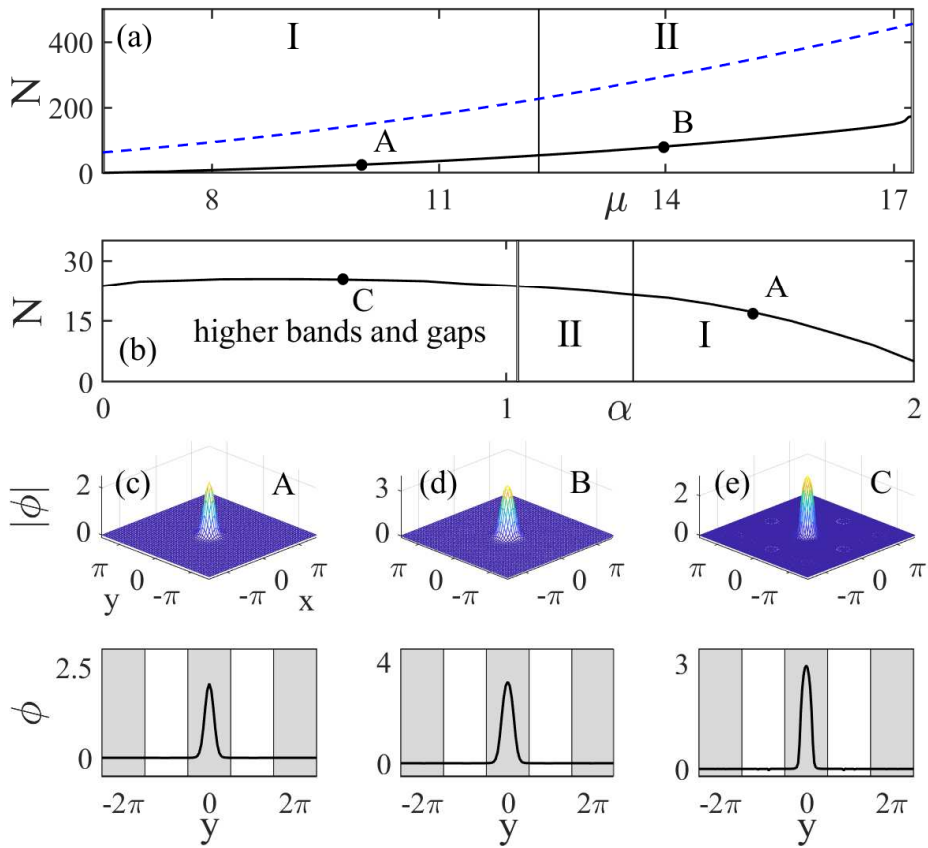}
\end{center}
\caption{Norm $N$ vs. chemical potential $\protect\mu $ (a) and L\'{e}vy
index $\protect\alpha $ (b) for 2D gap solitons in the system with the deep
optical lattice. The parameters are $\protect\alpha =1.6$ for (a) and $%
\protect\mu =10$ for (b). Panel (a) shows the results for the GSs in the two
lowest bandgaps, separated by the flat Bloch band at $\protect\mu \approx
12.3$. Results produced by the TF approximation, based on Eqs. (\protect\ref%
{TF-2D}) and (\protect\ref{NTF2D}), are shown by the blue dashed line. The
top and bottom plots in panels (c, d, e) display overall shapes and cross
sections of three typical solitons corresponding to points A, B, C,
severally, marked in panels (a) and (b). Points A, B, and C belong, respectively, to the first, second, and seventh finite bandgaps. Other parameters are:
(c) $\protect\mu =10$, $\protect\alpha =1.6$; (d) $\protect\mu =14$, $%
\protect\alpha =1.6$; (e) $\protect\mu =10$, $\protect\alpha =0.6$.}
\label{fig4}
\end{figure}


\begin{figure}[t]
\begin{center}
\includegraphics[width=0.7\columnwidth]{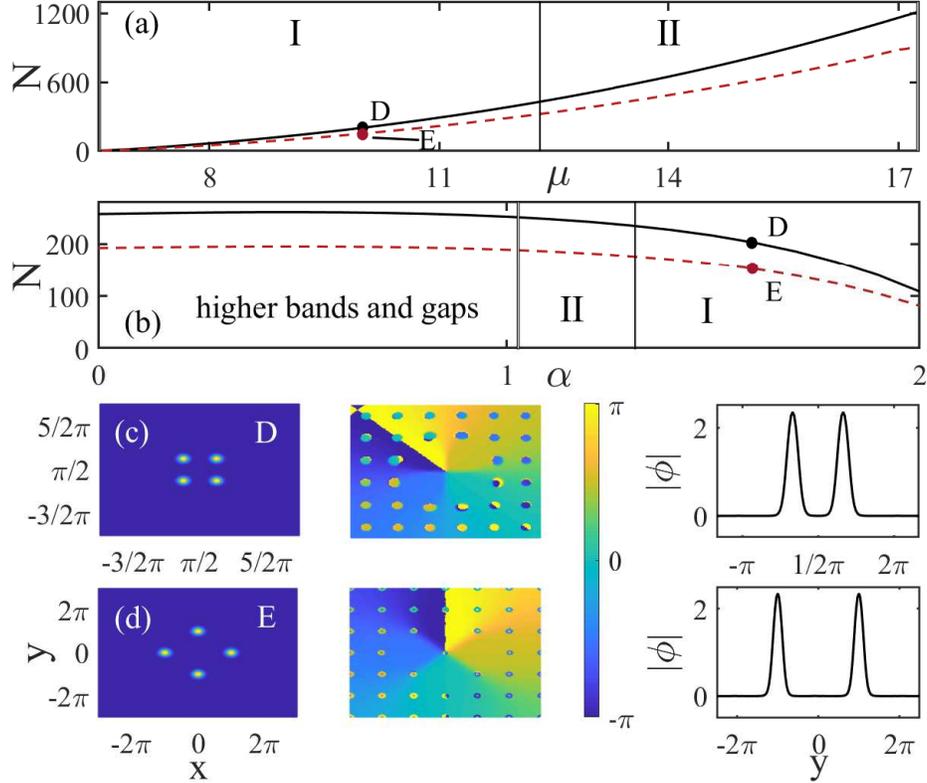}
\end{center}
\caption{The black and red lines show norm $N$ vs. chemical potential $%
\protect\mu $ (a) and L\'{e}vy index $\protect\alpha $ (b) for 2D gap
vortices of the \textquotedblleft square" and \textquotedblleft rhombus"
types, respectively, composed of four peaks carrying imprinted topological
charge $S=1$. The two species of the gap vortices, marked as points D and E
in (a,b), with equal values of the chemical potential and L\'{e}vy index ($%
\protect\alpha =1.6$, $\protect\mu =10$) are depicted in (c) and (d)
respectively. The corresponding phase structures and cross sections at $x=0$
are also displayed in (c) and (d). }
\label{fig5}
\end{figure}


\section{Two-dimensional gap solitons and solitary vortices}

\vspace{5mm}

To address 2D localized gap modes in the fractional setting under the
consideration, the corresponding bandgap structure for potential (\ref{2D})
should be produced at first. The \textquotedblleft egg-carton" shape of the
potential is displayed in Fig. \ref{fig3}(a), its first Brillouin zone for
the wave vector ($k_{x},k_{y}$) is shown, in the reciprocal lattice space,
in Fig. \ref{fig3}(c), and the resulting bandgap structure for the deep OL
with $V_{0}=40$ is exhibited in the plane of $\left( \alpha ,\mu \right) $
in Fig. \ref{fig3}(b), where the vertical red dashed line represents the
fixed value of LI, $\alpha =1.6$. In Fig. \ref{fig3}(b), higher-order
quasi-flat bands and gaps between them are not plotted, as we here focus on
the localized modes in the first two bandgaps. The study of modes in
higher-order gaps may be a subject for a separate work. We see in Fig. \ref%
{fig3}(d) that the 2D deep lattice, as well as its 1D counterpart, gives
rise to the flat-band spectrum, which also obeys scaling relation (\ref{mu}).

\begin{figure}[t]
\begin{center}
\includegraphics[width=0.95\columnwidth]{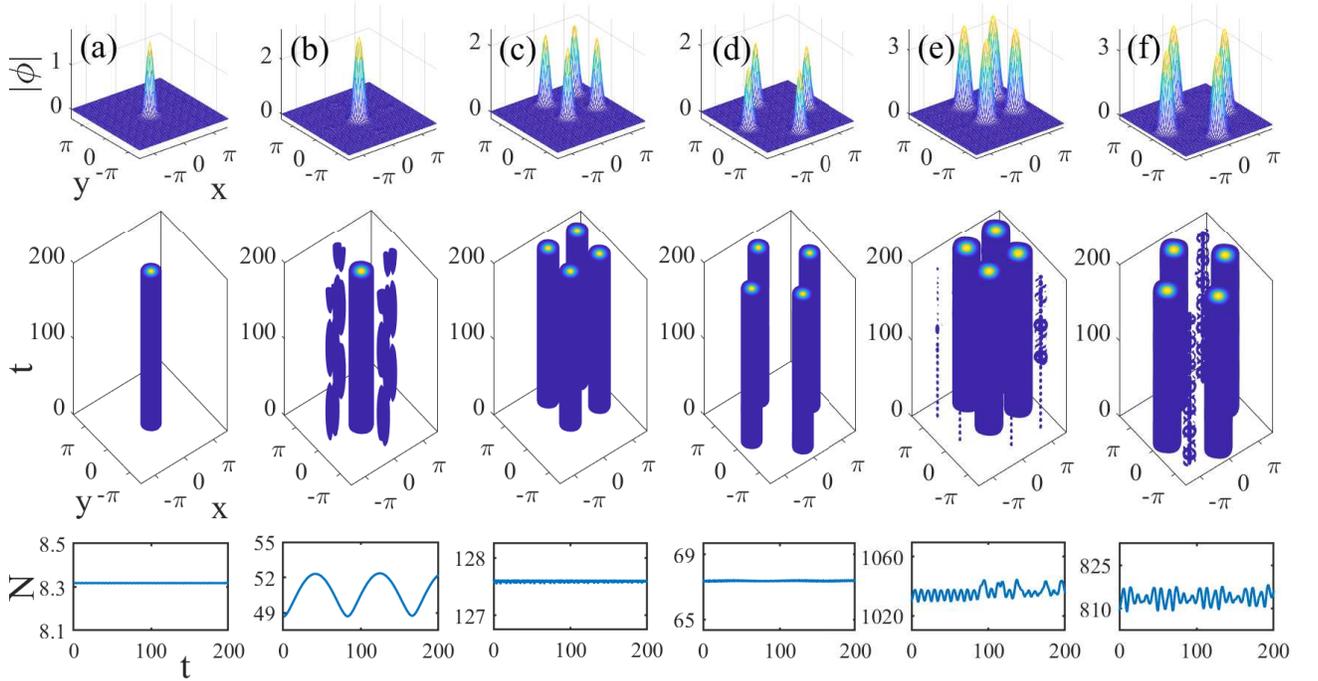}
\end{center}
\caption{Profiles of 2D GSs and gap vortices with topological charge $S=1$
in the fractional nonlinear system with the OL-potential depth $V_{0}=40$
(see Eq. (\protect\ref{2D})) and L\'{e}vy index $\protect\alpha =1.6$. The
corresponding perturbed evolution and norm $N$ vs. time are depicted in the
second and third lines, respectively.}
\label{fig6}
\end{figure}

\subsection{ Two-dimensional Fundamental GSs}

\vspace{5mm}

The simplest species of 2D gap modes represents fundamental GSs with a
single peak, similar to their 1D counterparts considered above. The family
of such GSs in the two lowest bandgaps is characterized by $N(\mu )$ and $%
\mu (\alpha )$ curves plotted in Figs. \ref{fig4}(a) and \ref{fig4}(b),
respectively. Note that the former curve agrees with the above-mentioned
anti-VK criterion, $dN/d\mu >0$. \vspace{5mm}

Two representative examples of the fundamental GSs, belonging to the first
and second finite bandgaps, are displayed in Figs. \ref{fig4}(c,e) and \ref%
{fig4}(d), respectively. In addition, their 1D cross-section along $x=0$ are
plotted in the bottom plots of Fig. \ref{fig4}, indicating that the 2D
fundamental GSs, as well as their 1D counterparts reported above, are
essentially confined to a single cell of the lattice, with virtually
invisible tails. In agreement with Eq. (\ref{tail}), the absence of
conspicuous tails is a specific feature of the fractional nonlinear systems
with the lattice potential and $\alpha <2$, the situation being different
for traditional GSs in the case of the normal diffraction ($\alpha =2$).
\vspace{5mm}

The TF approximation can be applied to the 2D fundamental GSs as well. It
yields%
\begin{equation}
\phi _{\mathrm{TF}}^{2}(x,y)=\left\{
\begin{array}{c}
\mu -V_{0}\left( \sin ^{2}x+\sin ^{2}y\right) ,~\mathrm{at}\ \sin ^{2}x+\sin
^{2}y<\mu /V_{0}, \\
0,~\mathrm{at}~\sin ^{2}x+\sin ^{2}y>\mu /V_{0}.%
\end{array}%
\right.  \label{TF-2D}
\end{equation}%
The respective norm was found numerically as%
\begin{equation}
N_{\mathrm{TF}}^{\mathrm{(2D)}}=\int \int \phi _{\mathrm{TF}}^{2}(x,y)dx,
\label{NTF2D}
\end{equation}%
where the integral should be calculated numerically over the area
corresponding to the top line in Eq. (\ref{TF-2D}). Figure \ref{fig4}(a)
shows that the norm predicted by this approximation is essentially different
from the numerically exact counterpart, hence we conclude that the results
for the 2D solitons are nontrivial, in the sense that they are essentially
affected by the fractional diffraction. \vspace{5mm}

\subsection{Vortex solitons}

The usual 2D nonlinear Schr\"{o}dinger equation with the square-shaped OL
potential (\ref{2D}) and self-defocusing cubic nonlinearity gives rise, in
addition to fundamental solitons, to ones with embedded integer vorticity $S$
\cite{HS}. Here we aim to construct localized gap vortices composed of four
peaks, onto which the overall phase circulation $2\pi $, corresponding to $%
S=1$, is imprinted. It is known that such patterns represent robust
gap-vortex modes in many physical settings with the usual (non-fractional)
diffraction \cite{FPC,RMP06,soliton-periodic,PC,NL-RMP,Frac-CQ2d,L-NL2d}.
Very recently, similar vortex solitons were found in a discrete fractional
model with the self-focusing nonlinearity (unlike the defocusing nonlinear
term considered here) \cite{FDNLS-vortex}. \vspace{5mm}

The four-peak configuration of the gap vortices can be arranged in the shape
of a densely packed square (alias an intersite-centered vortex), or a
loosely packed rhombus (an onsite-centered one), with an empty site at the
center \cite{PhysicaD}. Examples of these patterns are displayed in Figs. %
\ref{fig5}(c) and \ref{fig5}(d), respectively. Families of the vortex GSs,
with $S=1$, are characterized by the respective dependences $N(\mu )$ and $%
\mu (\alpha )$, which are shown in Figs. \ref{fig5}(a) for the
\textquotedblleft squares", and \ref{fig5}(b) for the \textquotedblleft
rhombuses". It is seen that the former species of the gap vortices has a
larger norm, which may be attributed to stronger interactions between the
four peaks, separated by an essentially smaller distance than in the
rhombuses, as seen in Figs. \ref{fig5}(c) and \ref{fig5}(d). \vspace{5mm}


\subsection{Dynamics of two-dimensional fundamental GSs and vortex solitons}

In Figs. \ref{fig6}(a) and \ref{fig6}(b), we show profiles and perturbed
evolution of a stable 2D fundamental GS existing in the first finite
bandgap, and of an unstable one found near the upper edge of the same
bandgap. It is seen that the GS, which was predicted to be stable by the
analysis of small perturbations, is indeed stable in direct simulations,
while the unstable one develops regular oscillations. In the bottom line of
both panels, the dependence of the solitons' norm $N$ on time corroborates
these conclusions. It is seen that the instability transforms the 2D soliton
into a breather, but does not destroy it. \vspace{5mm}

Typical profiles of two kinds of vortex GSs, i.e., the above-mentioned
squared and vortices, are plotted in Figs. \ref{fig6}(c-f). They display
stable and unstable gap vortices, respectively, in the first finite bandgap
and near the upper edge of the same bandgap. The corresponding perturbed
evolution and dependence of the norm on time are depicted in the second and
third lines of the panels. It is seen, in particular, that unstable vortices
develop random oscillations, but survive as topologically organized patterns.
We stress that, being consistent with the 1D case, the 2D GSs and vortices are exceptionally stable, except for near edges of the bandgaps.

\section{Conclusion and discussion}

We have presented the analysis of the existence, structure, and dynamics of
localized gap modes, including 1D and 2D fundamental GSs (gap solitons), as
well as 2D gap vortices, in the fractional system including the deep OL
(optical-lattice) potential and self-repulsive nonlinearity. The system's
spectrum features finite bandgaps separated by nearly flat bands. Fractional
solitons in such deep periodic potentials were not studies previously, and,
unlike the model with the normal (non-fractional) diffraction, they cannot
be obtained in the discrete approximation. We have found that the 1D GSs,
belonging to the first and second finite bandgaps of the underlying spectral
structure, are strongly confined in a single lattice cell, on the contrary
to multi-cell GSs supported by shallow OLs. The linear-stability analysis
and direct simulations of the perturbed evolution have identified stability
and instability regions for the 1D GSs. They are stable inside the first and
second finite bandgaps and unstable in narrow regions near the gap edges.
The 2D GSs and vortices found here are, generally, stable too, being
unstable only very close to edges of the bandgaps. The predicted localized
modes may be observed in experiments with BEC loaded into deep OLs, as well
as in optical waveguides composed of alternating layers with large and small
values of the refractive index. \vspace{5mm}

It may be interesting to develop the analysis for vortex GSs with higher
values of the winding number, $S\geq 2$. This work may be extended for
physical systems modeled by the NLFSEs (nonlinear fractional Schr\"{o}dinger
equation) for temporal optical GSs \cite{GS-TNLBW}. It will also be natural
to consider possible GSs in bimodal fractional models, such as those based
on dual-core optical couplers \cite{Frac-coupler}.

\medskip \textbf{Conflict of Interest}

The authors declare no conflicts of interest.

\medskip \textbf{Acknowledgements}

This work was supported by the National Natural Science Foundation of China
(NSFC) (Nos. 61690224, 61690222,12074423); Israel Science Foundation (grant
No. 1286/17).


\begin{thebibliography}{99}

\bibitem{FC-book0} A. A. Kilbas, H. M. Srivastava, and J. J. Trujillo,
\textit{Theory and Applications of Fractional Differential Equations}
(Elsevier, 2006).

\bibitem{FC-book1} V. E. Tarasov, \textit{Fractional Dynamics: Applications
of Fractional Calculus to Dynamics of Particles, Fields and Media},
(Springer, 2010).

\bibitem{FC-book2} I. Petr\'{a}\v{s}, \textit{Fractional-Order Nonlinear
Systems: Modeling, Analysis and Simulation}, (Springer, 2010).

\bibitem{FC-book3} J. Klafter, S. C. Lim, and R. Metzler, \textit{Fractional
Dynamics: recent advances}, (World Scientific, 2012).

\bibitem{FC-book4} R. Herrmann, \textit{Fractional Calculus: An Introduction
for Physicists}, 2nd ed. (World Scientific, 2014).

\bibitem{Lask1} N. Laskin, Fractional quantum mechanics and L\'{e}vy path
integrals, \textit{Phys. Lett. A} \textbf{268}, 298-305 (2000).

\bibitem{Lask2} N. Laskin, Fractional quantum mechanics, \textit{Phys. Rev. E%
} \textbf{62}, 3135 (2000).

\bibitem{Lask3} N. Laskin, Fractional Schr\"{o}dinger equation, \textit{%
Phys. Rev. E} \textbf{66}, 056108 (2002).

\bibitem{Fujioka} J. Fujioka, A. Espinosa, and R. F. Rodriguez, Fractional
optical solitons, \textit{Phys. Lett. A} \textbf{374}, 1126-1134 (2010).

\bibitem{Frac-optics} S. Longhi, Fractional Schr\"{o}dinger equation in
optics, \textit{Opt. Lett.} \textbf{40}, 1117-1120 (2015).

\bibitem{frac-BEC-diffusion-E} Y. Sagi, M. Brook, I. Almog and N. Davidson,
Observation of anomalous diffusion and fractional self-similarity in one
dimension, \textit{Phys. Rev. Lett.} \textbf{108}, 093002 (2012).

\bibitem{frac-BEC-diffusion-T} A. D. Kessler and E. Barkai, Theory of
fractional L\'{e}vy kinetics for cold atoms diffusing in optical lattices,
\textit{Phys. Rev. Lett.} \textbf{108}, 230602 (2012).

\bibitem{Frac-condensed} B. A. Stickler, Potential condensed-matter
realization of space-fractional quantum mechanics: The one-dimensional L\'{e}%
vy crystal, \textit{Phys. Rev. E} \textbf{88}, 012120 (2013).

\bibitem{Frac-polariton} F. Pinsker, W. Bao, Y. Zhang, H. Ohadi, A.
Dreismann, and J. J. Baumberg, Fractional quantum mechanics in polariton
condensates with velocity-dependent mass, \textit{Phys. Rev. B} \textbf{92},
195310 (2015).

\bibitem{Frac-book} \textit{N. Laskin, Fractional quantum mechanics}, (World
Scientific, 2018).

\bibitem{Frac1} Y. Zhang, X. Liu, M. R. Beli\'{c}, W. Zhong, Y. Zhang and M.
Xiao, Propagation dynamics of a light beam in a fractional Schr\"{o}dinger
equation, \textit{Phys. Rev. Lett.} \textbf{115}, 180403 (2015).

\bibitem{Frac2} Y. Zhang, H. Zhong, M. R. Beli\'{c}, Y. Zhu, W. P. Zhong, Y.
P. Zhang, D. N. Christodoulides, and M. Xiao, $\mathcal{PT}$ symmetry in a
fractional Schr\"{o}dinger equation, \textit{Laser Photon. Rev.} \textbf{10}%
, 526 (2016).

\bibitem{Frac3} W. P. Zhong, M. R. Beli\'{c}, B. A. Malomed, Y. Zhang, and
T. Huang, Spatiotemporal accessible solitons in fractional dimensions,
\textit{Phys. Rev. E} \textbf{94}, 012216 (2016).

\bibitem{Frac3b} D. Zhang, Y. Zhang, Z. Zhang, N. Ahmed, Y. Zhang, F. Li, M.
R. Belic, and M. Xiao, Unveiling the link between fractional Schr\"{o}dinger
equation and light propagation in honeycomb lattice, \textit{Ann. Phys.
(Berlin)} \textbf{529}, 1700149 (2017).

\bibitem{Frac4} L. Zhang, C. Li, H. Zhong, C. Xu, D. Lei, Y. Li, and D. Fan,
Propagation dynamics of super-Gaussian beams in fractional Schr\"{o}dinger
equation: From linear to nonlinear regimes, \textit{Opt. Express} \textbf{24}%
, 14406 (2016).

\bibitem{Frac5} L. Zhang, Z. He, C. Conti, Z. Wang, Y. Hu, D. Lei, Y. Li,
and D. Fan, Modulational instability in fractional nonlinear Schr\"{o}dinger
equation, \textit{Commun. Nonlinear Sci. Numer. Simulat.} \textbf{48}, 531
(2017).

\bibitem{Frac5b} L. Zhang, X. Zhang, H. Wu, C. Li, D. Pierangeli ,and D.
Fan, Anomalous interaction of Airy beams in the fractional nonlinear Schr%
\"{o}dinger equation, \textit{Opt. Express} \textbf{27}, 27936 (2019).

\bibitem{Frac6} C. Huang and L. Dong, Gap solitons in the nonlinear
fractional Schr\"{o}dinger equation with an optical lattice, \textit{Opt.
Lett.} \textbf{41}, 5636 (2016).

\bibitem{Frac7} X. Yao and X. Liu, Off-site and on-site vortex solitons in
space-fractional photonic lattices, \textit{Opt. Lett.} \textbf{43}, 5749
(2018).

\bibitem{Frac7b} X. Yao and X. Liu, Solitons in the fractional Schr\"{o}%
dinger equation with parity-time-symmetric lattice potential, \textit{%
Photonics Res.} \textbf{6}, 875-879 (2018).

\bibitem{Dong} J. Xiao, Z. Tian, C. Huang, and L. Dong, Surface gap solitons
in a nonlinear fractional Schr\"{o}dinger equation, \textit{Opt. Express}
\textbf{26}, 2650-2658 (2018).

\bibitem{Frac8} M. Chen, S. Zeng, D. Lu, W. Hu, and Q. Guo, Optical
solitons, self-focusing, and wave collapse in a space-fractional Schr\"{o}%
dinger equation with a Kerr-type nonlinearity, \textit{Phys. Rev. E} \textbf{%
98}, 022211 (2018).

\bibitem{Frac9} M. Chen, Q. Guo, D. Lu, and W. Hu, Variational approach for
breathers in a nonlinear fractional Schr\"{o}dinger equation, \textit{%
Commun. Nonlinear Sci. Numer. Simulat.} \textbf{71}, 73 (2019).

\bibitem{Frac10} L. Zeng and J. Zeng, One-dimensional solitons in fractional
Schr\"{o}dinger equation with a spatially periodical modulated nonlinearity:
nonlinear lattice, \textit{Opt. Lett.} \textbf{44}, 2661 (2019).

\bibitem{Frac11} L. Dong, C. Huang, and W. Qi, Nonlocal solitons in
fractional dimensions, \textit{Opt. Lett.} \textbf{44}, 4917 (2019).

\bibitem{Frac12} J. Xie, X. Zhu, and Y. He, Vector solitons in nonlinear
fractional Schr\"{o}dinger equations with parity-time-symmetric optical
lattices, \textit{Nonlinear Dyn.} \textbf{97}, 1287 (2019).

\bibitem{Frac-CQ1d} L. Zeng and J. Zeng, One-dimensional gap solitons in
quintic and cubic-quintic fractional nonlinear Schr\"{o}dinger equations
with a periodically modulated linear potential, \textit{Nonlinear Dyn.}
\textbf{98}, 985 (2019).

\bibitem{Frac-CQ2d} L. Zeng and J. Zeng, Preventing critical collapse of
higher-order solitons by tailoring unconventional optical diffraction and
nonlinearities, \textit{Commun. Phys.} \textbf{3}, 26 (2020).

\bibitem{Frac-SSB} J. Chen and J. Zeng, Spontaneous symmetry breaking in
purely nonlinear fractional systems, \textit{Chaos} \textbf{30}, 063131
(2020).

\bibitem{Frac-saturable} J. Shi and J. Zeng, 1D Solitons in Saturable
Nonlinear Media with Space Fractional Derivatives, \textit{Ann. Phys.
(Berlin)} \textbf{532}, 1900385 (2020).

\bibitem{Frac-coupler} L. Zeng and J. Zeng, Fractional quantum couplers,
\textit{Chaos, Solitons and Fract.} \textbf{140}, 110271 (2020).

\bibitem{Frac-SSBLinear} P. Li, B. A. Malomed, and D. Mihalache, Symmetry
breaking of spatial Kerr solitons in fractional dimension, \textit{Chaos,
Solitons and Fract.} \textbf{132}, 109602 (2020).

\bibitem{Frac-CQ} P. Li, B. A. Malomed, and D. Mihalache, Vortex solitons in
fractional nonlinear Schr\"odinger equation with the cubic-quintic
nonlinearity, \textit{Chaos, Solitons and Fract.} \textbf{137}, 109783
(2020).

\bibitem{Frac-PRR} D. Colas, Self-accelerating beam dynamics in the space
fractional Schr\"{o}dinger equation, \textit{Phys. Rev. Research} \textbf{2}%
, 033274 (2020).

\bibitem{Frac-CGL} Y. Qiu, B. A. Malomed, D. Mihalache, X. Zhu, L. Zhang,
and Y. He, Soliton dynamics in a fractional complex Ginzburg-Landau model,
\textit{Chaos, Solitons Fract} \textbf{131}, 109471 (2020).

\bibitem{Frac-GS-PT} L. Li, H.-G. Li, W. Ruan, F.-C. Leng, and X.-B. Luo,
Gap solitons in parity-time-symmetric lattices with fractional-order
diffraction, \textit{J. Opt. Soc. Am. B} 37, 488-494 (2020).

\bibitem{Frac-PTOL} X. Zhu, F. Yang, S. Cao, J. Xie, and Y. He, Multipole
gap solitons in fractional Schr\"{o}dinger equation with
parity-time-symmetric optical lattices, \textit{Opt. Express} \textbf{28},
1631-9 (2020).

\bibitem{Pengfei} P. Li, R. Li, and C. Dai, Existence, symmetry breaking
bifurcation and stability of two-dimensional optical solitons supported by
fractional diffraction, Opt. Exp. \textbf{29}, 3193-3210 (2021).

\bibitem{review} B. A. Malomed, Optical solitons and vortices in fractional
media: A mini-review of recent results, \textit{Photonics} \textbf{8}, 353
(2021).

\bibitem{FPC} Y. S. Kivshar and G. P. Agrawal, \textit{Optical Solitons:
From Fibers to Photonic Crystals} (Academic, 2003).

\bibitem{RMP06} O. Morsch and M. Oberthaler, Dynamics of Bose-Einstein
condensates in optical lattices, \textit{Rev. Mod. Phys.} \textbf{78}, 179
(2006).

\bibitem{soliton-periodic} B. A. Malomed, \textit{Soliton Management in
Periodic Systems} (Springer, 2006).

\bibitem{PC} J. D. Joannopoulos, S. G. Johnson, J. N. Winn, and R. D. Meade,
\textit{Photonic Crystals: Molding the Flow of Light }(Princeton University
Press, 2008).

\bibitem{NL-RMP} Y. V. Kartashov, B. A. Malomed, and L. Torner, Solitons in
nonlinear lattices, \textit{Rev. Mod. Phys.} \textbf{83}, 247-306 (2011).

\bibitem{PL} I. L. Garanovich, S. Longhi, A. A. Sukhorukov, and Y. S.
Kivshar, Light propagation and localization in modulated photonic lattices
and waveguides, \textit{Phys. Rep. } \textbf{518}, 1-79 (2012).

\bibitem{NRP} Y. V. Kartashov, G. E. Astrakharchik, B. A. Malomed, and L.
Torner, Frontiers in multidimensional self-trapping of nonlinear fields and
matter, \textit{Nat. Rev. Phys. } \textbf{1}, 185-197 (2019).

\bibitem{BEC-darkGap} L. Zeng and J. Zeng, Gap-type dark localized modes in
a Bose-Einstein condensate with optical lattices, \textit{Adv. Photonics}
\textbf{1}, 046006 (2019).

\bibitem{NL-focus} J. Shi and J. Zeng, Self-trapped spatially localized
states in combined linear-nonlinear periodic potentials, \textit{Front. Phys.%
} \textbf{15}, 12602 (2020).

\bibitem{BEC-darkGapQ} J. Li and J. Zeng, Dark matter-wave gap solitons in
dense ultracold atoms trapped by a one-dimensional optical lattice, \textit{%
Phys. Rev. A }\textbf{103}, 013320 (2021).

\bibitem{BEC-EIT1d} Z. Chen and J. Zeng, Localized gap modes of coherently
trapped atoms in an optical lattice, \textit{Opt. Express }\textbf{29}, 3011
(2021).

\bibitem{BEC-EIT2d} Z. Chen and J. Zeng, Two-dimensional optical gap
solitons and vortices in a coherent atomic ensemble loaded on optical
lattices, \textit{Commun. Nonlinear Sci. Numer. Simulat.} \textbf{102},
105911 (2021).

\bibitem{GS-FBG} B. J. Eggleton, R. E.~Slusher, C. M.~de Sterke, P. A.~Krug,
and J. E.~Sipe, Bragg grating solitons, \textit{Phys. Rev. Lett.} \textbf{76}%
, 1627 (1996).

\bibitem{GS-WA} D. Mandelik, R. Morandotti, J. S. Aitchison, and Y.
Silberberg, Gap solitons in waveguide arrays, \textit{Phys. Rev. Lett.}
\textbf{92}, 093904 (2004).

\bibitem{moving-gap-sol} J. T. Mok, C. M. de Sterke, I. C. M. Litte, and B.
J. Eggleton, Dispersionless slow light using gap solitons,\ \textit{Nature
Phys}. \textbf{2}, 775-780 (2006).

\bibitem{GS-HPL} O. Peleg, G. Bartal, B. Freedman, O. Manela, M. Segev, and
D. N. Christodoulides, Conical diffraction and gap solitons in honeycomb
photonic lattices, \textit{Phys. Rev. Lett.} \textbf{98}, 103901 (2007).

\bibitem{GS-BEC} B. Eiermann, T. Anker, M. Albiez, M. Taglieber, P.
Treutlein, K. P. Marzlin and M. K. Oberthaler, Bright Bose-Einstein gap
solitons of atoms with repulsive interaction, \textit{Phys. Rev. Lett.}
\textbf{92}, 230401 (2004).

\bibitem{PhotCryst} J. D. S. Joannopoulos, J. N. Winn, and R. D. Meade,
\textit{Photonic Crystals: Molding the Flow of Light} (Princeton University
Press, Princeton, 2008).

\bibitem{GS-EPC1} E. A. Ostrovskaya, J. Abdullaev, M. D. Fraser, A. S.
Desyatnikov, and Y. S. Kivshar, Self-localization of polariton condensates
in periodic potentials, \textit{Phys. Rev. Lett.} \textbf{110}, 170407
(2013).

\bibitem{GS-EPC2} E. A. Cerda-M\'{e}ndez, D. Sarkar, D. N. Krizhanovskii, S.
S. Gavrilov, K. Biermann, M. S. Skolnick, and P. V. Santos,
Exciton-polariton gap solitons in two-dimensional lattices, \textit{Phys.
Rev. Lett. }\textbf{111}, 146401 (2013).

\bibitem{GS-EPC3} D. Tanese, H. Flayac, D. Solnyshkov, A. Amo, A. Lemaitre,
E. Galopin, R. Braive, P. Senellart, I. Sagnes, G. Malpuech, and J. Bloch,
Polariton condensation in solitonic gap states in a one-dimensional periodic
potential, \textit{Nat. Commun.} \textbf{4}, 1749 (2013).

\bibitem{moire1} P. Wang, Y. Zheng, X. Chen, C. Huang, Y. V. Kartashov, L.
Torner, V. V. Konotop, and F. Ye, Localization and delocalization of light
in photonic moir\'{e} lattices, \textit{Nature (London)} \textbf{577}, 422
(2020).

\bibitem{moire2} Q. Fu, P. Wang, C. Huang , Y. V. Kartashov, L. Torner, V.
V. Konotop, and F. Ye, Optical soliton formation controlled by angle
twisting in photonic moir\'{e} lattices, \textit{Nature Photonics} \textbf{14%
}, 663(2020).

\bibitem{Smerzi} A. Trombettoni and A. Smerzi, Discrete solitons and
breathers with dilute Bose-Einstein condensates, \textit{Phys. Rev. Lett.}
\textbf{86}, 2353-2356 (2001).

\bibitem{staggered} D. Cai, A. R. Bishop, and N. Gr\o nbech-Jensen,
Localized states in discrete nonlinear Schr\"{o}dinger equations, \textit{%
Phys. Rev. Lett}. \textbf{72}, 591-595 (1994).


\bibitem{Discrete1d} M. I. Molina, The fractional discrete nonlinear Schr%
\"{o}dinger equation, \textit{Phys. Lett. A} \textbf{384}, 126180 (2020).

\bibitem{Discrete2d} M. I. Molina, The two-dimensional fractional discrete
nonlinear Schr\"{o}dinger equation, \textit{Phys. Lett. A} \textbf{384},
126835 (2020).

\bibitem{FDNLS-vortex} C. M.-Cort\'{e}s, and M. I. Molina, Fractional
discrete vortex solitons, \textit{Opt. Lett. }\textbf{46}, 2256 (2021).

\bibitem{HS} H. Sakaguchi and B. A. Malomed, Two-dimensional loosely and
tightly bound solitons in optical lattices and inverted traps, \textit{J.
Phys. B: At. Mol. Opt. Phys.} \textbf{37}, 2225-2239 (2004).

\bibitem{Pit} L. P. Pitaevskii and S. Stringari, \textit{Bose-Einstein
Condensation} (Oxford University Press, Oxford, 2003).

\bibitem{Salasnich} L. Salasnich, A. Parola, and L. Reatto, Effective wave
equations for the dynamics of cigar-shaped and disk-shaped Bose condensates,
\textit{Phys. Rev. A} 65, 043614 (2002).

\bibitem{Riesz} M. Cai and C. P. Li, On Riesz derivative, \textit{Fractional
Calculus and Applied Analysis} \textbf{22}, 287-301 (2019).

\bibitem{GW-BEC} Th. Anker, M. Albiez, R. Gati, S. Hunsmann, B. Eiermann, A.
Trombettoni and M. K. Oberthaler, Nonlinear self-trapping of matter waves in
periodic potentials, \textit{Phys. Rev. Lett.} \textbf{94}, 020403 (2005).

\bibitem{MSOM} J. Yang, \textit{Nonlinear Waves in Integrable and
Nonintegrable Systems}, (SIAM: Philadelphia, 2010).

\bibitem{flat0} E. J. Bergholtz and Z. Liu, Topological flat band models and
fractional Chern insulators, Int. J. Mod. Phys. B \textbf{27}, 1330017
(2013).

\bibitem{flat1} A. Amo and J. Bloch, Exciton-polaritons in lattices: A
non-linear photonic simulator, \textit{Comp. Rend. Phys}. \textbf{17},
934-945 (2016).

\bibitem{Flach} D. Leykam, A. Andreanov, and S. Flach, Artificial flat band
systems: from lattice models to experiments, \textit{Adv. Phys. X} \textbf{3}%
, 1-25 (2018).

\bibitem{flat2} S. Q. Xia, L. Q. Tang, S. Q. Xia, J. N. Ma, W. C. Yan, D. H.
Song, Y. Hu, J. J. Xu, and Z. G. Chen, Novel phenomena in flatband photonic
structures: from localized states to real-space topology, \textit{Acta
Physica Sinica} \textbf{69}, 154207 (2020).

\bibitem{flat} S. Q. Xia, D. H. Song, N. Wang, X. Y. Liu, J. N. Ma, L. Q.
Tang, H. Buljan, and Z. G. Chen, Topological phenomena demonstrated in
photorefractive photonic lattices [Invited], \textit{Opt. Materials Exp}.
\textbf{11}, 1292-1312 (2021).

\bibitem{no exact solution} M. Jeng, S. L. Y. Xu, E. Hawkins, and J. M.
Schwarz, On the nonlocality of the fractional Schr\"{o}dinger equation,
\textit{J. Math. Phys}. 51, 062102 (2010).

\bibitem{L-NL1d} H. Sakaguchi and B. A. Malomed, Solitons in combined linear
and nonlinear lattice potentials, \textit{Phys. Rev. A} \textbf{81}, 013624
(2010).

\bibitem{L-NL2d} J. Zeng and B. A. Malomed, Two-dimensional solitons and
vortices in media with incommensurate linear and nonlinear lattice
potentials, \textit{Phys. Scr. } \textbf{T149}, 014035 (2012).

\bibitem{GS-PT1} J. Zeng and Y. Lan, Two-dimensional solitons in $\mathcal{PT%
}$ linear lattice potentials, \textit{Phys. Rev. E }\textbf{85}, 047601
(2012).

\bibitem{VK} M. Vakhitov and A. Kolokolov, Stationary solutions of the wave
equation in a medium with nonlinearity saturation, \textit{Radiophys.
Quantum Electron.} \textbf{16}, 783 (1973).

\bibitem{Berge} L. Berg\'{e}, Wave collapse in physics: principles and
applications to light and plasma waves, \textit{Phys. Rep}. \textbf{303},
259-372 (1998).

\bibitem{Fibich} G. Fibich, \textit{The Nonlinear Schr\"{o}dinger equation:
Singular Solutions and Optical Collapse} (Springer: Cham, Heidelberg, New
York, Dordrecht, London, 2015).


\bibitem{PhysicaD} B. A. Malomed, (INVITED)\ Vortex solitons: Old results
and new perspectives, \textit{Physica D} \textbf{399}, 108-137 (2019).

%
%

\bibitem{GS-TNLBW} C. Bersch, G. Onishchukov, and U. Peschel, Optical gap
solitons and truncated nonlinear Bloch waves in temporal lattices, \textit{%
Phys. Rev. Lett.} \textbf{109}, 093903 (2012).
\end{thebibliography}
\end{document}